\documentclass[
    ,final            
  ]
  {aipproc}

\layoutstyle{6x9}

\usepackage{epsfig}
\usepackage{wrapfig,rotating}
\usepackage{amssymb}
\usepackage{amsmath}
\newcommand{\be}{\begin{equation}}
\newcommand{\ee}{\end{equation}}
\newcommand{\beq}{\begin{equation}}
\newcommand{\eeq}{\end{equation}}
\newcommand{\bea}{\begin{eqnarray}}
\newcommand{\eea}{\end{eqnarray}}

\newcommand{\rr}{\mbox{\boldmath $r$}}

\begin{document}

\title{Deeply Virtual Compton Scattering at HERA and perspectives at CERN}


\author{Laurent Schoeffel}{
  address={CEA Saclay, DAPNIA-SPP, 91191 Gif-sur-Yvette Cedex, France}
}

\begin{abstract}
 Standard parton distribution functions contain neither information on the
correlations between partons nor on their transverse motion,
then a vital knowledge about the three dimensional 
structure of the nucleon is lost.
Hard exclusive processes, in particular DVCS, are essential reactions to go beyond
this standard picture.
In the following, we examine the most recent data in view of the dipole
model predictions  and their implication
on the quarks/gluons imaging (tomography) of the nucleon.

\end{abstract}

\maketitle


\section{Introduction}

Measurements of the deep-inelastic scattering (DIS) of leptons and nucleons, $e+p\to e+X$,
allow the extraction of Parton Distribution Functions (PDFs) which describe
the longitudinal momentum carried by the quarks, anti-quarks and gluons that
make up the fast-moving nucleons. 
These functions have been measured over a wide
kinematic range in the Bjorken scaling variable $x_{Bj}$ and
the photon virtuality $Q^2$.
While PDFs provide crucial input to
perturbative Quantum Chromodynamic (QCD) calculations of processes involving
hadrons, they do not provide a complete picture of the partonic structure of
nucleons. 
In particular, PDFs contain neither information on the
correlations between partons nor on their transverse motion,
then a vital knowledge about the three dimensional 
structure of the nucleon is lost.
Hard exclusive processes, in  which the
nucleon remains intact, have emerged in recent years as prime candidates to complement
this essentially one dimentional picture \cite{lolopic}. 

Indeed, within the investigation of hard exclusive
reactions in the Bjorken limit, the probe provided by the photon works as a clean tool reactions
in order to extract reliable knowledge on the substructure of strongly
interacting particles complementar to exclusive process.

The simplest exclusive process  is the deeply virtual
Compton scattering (DVCS) or exclusive production of real photon, $e + p \rightarrow e + \gamma + p$.
This process is of particular interest as it has both a clear
experimental signature and is calculable in perturbative QCD. 
The DVCS reaction can be regarded as the elastic scattering of the
virtual photon off the proton via a colourless exchange, producing a real photon in the final state  \cite{lolopic}. 
The recent measurements from the DESY $ep$ collider HERA 
at low $x_{Bj}$ ($x_{Bj}<0.01$) and large $Q^2$ (above 2 GeV$^2$) 
 are thus a decisive experimental step forward  \cite{dvcsh1,dvcszeus}. In the Bjorken scaling 
regime, 
QCD calculations assume that the exchange involves two partons, having
different longitudinal and transverse momenta, in a colourless
configuration. These unequal momenta or skewing are a consequence of the mass
difference between the incoming virtual photon and the outgoing real
photon. This skewedness effect can
 be interpreted in the context of generalised
parton distributions (GPDs) \cite{diehl,freund2,buk} or in the dipole model approach \cite{robi,marquet}.

A considerable interest of the high energy
situation at HERA is that it gives the important opportunity to constrain the gluon
contribution to GPDs and to study the evolution in virtuality $Q^2$ of the
quark and gluon distributions. On the other hand, recently the color
dipole formalism has provided a simultaneous description of photon
induced process. The inclusive deep inelastic reaction and the photon
diffractive dissociation has been successfully described and the study of
other exclusive processes such as DVCS is an important test of the color dipole
approach.

In the following, we examine the most recent data in view of the dipole
model predictions \cite{robi,marquet} and their implication
on the quarks/gluons imaging of the nucleon \cite{dvcsh1,lolotom,pire}.

\section{The color dipole model}

The colour dipole model  provides a simple 
unified picture of inclusive and diffractive processes and enables hard
and soft physics to be incorporated in a single dynamical framework. 
At high energies, in the proton's rest frame, the virtual photon fluctuates 
into a hadronic system (the simplest of which is a $q {\bar q}$ dipole) a 
long distance upstream of the target proton. The formation time of this hadronic 
system, and of the subsequent formation of the hadronic final state, is much longer 
than the interaction time with the target. 

DVCS is a good probe of the transition between soft and hard regimes in 
the dipole model for two reasons. Indeed, the transverse photon 
wave function can select large dipoles, even for large $Q^{2}$, and certainly 
for the $Q^2$ range $2 < Q^2 < 20$ GeV$^2$. Also, because the final photon is real,
DVCS is more sensitive to large dipoles than DIS at the same $Q^2$.

Then, in the colour dipole approach, the DIS (or DVCS) process can be seen as a
succession in time of three factorisable subprocesses: i)  the virtual
photon fluctuates in a quark-antiquark pair, ii) this colour dipole
interacts with the proton target, iii) the quark pair annihilates into a
virtual (or real) photon. The imaginary part of the DIS (or DVCS)
amplitude at $t=0$ is expressed in the simple way \cite{robi,marquet}
\begin{eqnarray} 
 {\cal I}m\, {\cal A}\,(W,Q_1,Q_2)  =  \sum_{T,L}\int \limits_0^1 dz \! 
 \int\limits_{0}^ {\infty} d^2\rr\, 
 \Psi_{T,L}^*(z,\,\rr,\,Q_1^2)\,\sigma_{dip}\,(z,\rr)\Psi_{T,L}
 (z,\,\rr,\,Q_2^2)\label{dvcsdip}\,,
\end{eqnarray} 
where $\Psi(z,\rr,Q_{1,2})$ are the light cone photon wave functions for
transverse and longitudinal photons. The quantity $Q_1$ is the virtuality
of the incoming photon, whereas $Q_2$ is the virtuality of the outgoing
photon. 
In the DIS case, one has $Q_1^2=Q_2^2=Q^2$ and for DVCS,
$Q_1^2=Q^2$ and $Q_2^2=0$. 
The relative transverse quark pair (dipole) separation is labeled by 
$\rr$ whilst $z$ (respec.\ $1-z$) labels the quark (antiquark)
longitudinal momentum fraction.

It should be noticed that the non-forward kinematics for DVCS is encoded in
the colour dipole approach through the different weight coming from the
photon wavefunctions in Eq. (\ref{dvcsdip}). The off-diagonal effects,
which affect the gluon and quark distributions in GPDs models,
should be included in the parameterisation of the dipole cross section. At
the present stage of the development of the  dipole formalism, we
have no accurate theoretical arguments on how to compute skewedness effects
from first principles. A consistent approach would be to compute the
scattering amplitude in the non-forward case, since the non-forward photon
wave function has been recently obtained.
In this case, the dipole cross section,
$\sigma_{dip}\,(x_1,x_2,\rr,\vec{\Delta })$, depends on the 
momenta $x_1$ and $x_2$ carried by the exchanged gluons, respectively, and
on the total transverse momentum transfer $\vec{\Delta}$. In this case,
additional information about the dependence upon $\vec{\Delta}$ is needed for
the QCD Pomeron and proton impact factor.  A first attempt in this direction
is done in \cite{robi}.

\section{Geometric Scaling}
\label{geom}
%
At very small values of the Bjorken scaling variable $x$
the saturation regime of QCD can be reached.
In this domain, the gluon density in the proton is so large that 
non-linear effects like gluon recombination tame its growth.
In the dipole model approach, the transition to the saturation regime is characterised by the so-called 
saturation scale parametrised here as $Q_s(x)=Q_0 ({{x_0}/{x}})^{-\lambda/2}$, where $Q_0$, $x_0$ and $\lambda$ are parameters. 
The transition to saturation occurs when $Q$ becomes comparable to 
$Q_s(x)$.
An important feature of dipole models that incorporate saturation is that the total cross section can be expressed as a function of the single variable $\tau$:

\begin{equation}
\sigma_{tot}^{\gamma^\ast p}(x,Q^2)  = \sigma_{tot}^{\gamma^\ast p}( \tau ) , \;\;  \mbox{with} \; \ \ \ \tau=\frac{Q^2}{Q_s^2(x)}.
\label{eq:tau}
\end{equation}
This property, called geometric scaling, has already been observed  to hold
for the total $ep$ DIS cross section \cite{golec} and in diffractive processes \cite{marquet}. 
It has also recently been addressed in the context of
exclusive processes including DVCS and extended to cases with non-zero momentum transfer to the proton \cite{robi}.
It is therefore interesting to test if the present DVCS measurements obey the geometric scaling laws predicted by such models.

\section{Latest News from the experimental Front}

\begin{figure}[!] 
    \includegraphics[width=8.0cm]{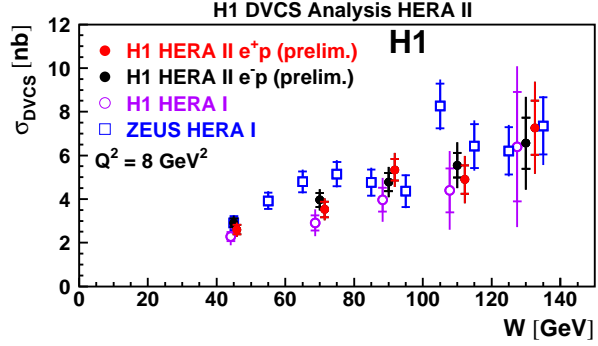}
  \caption{DVCS cross section for positrons/electrons samples as a function of
$W$ for  $Q^2=8$~GeV$^2$. 
}
\label{fig1}  
\end{figure}

Measurements of DVCS cross section have been realised  at HERA within the H1 and
ZEUS experiments \cite{dvcsh1,dvcszeus}. As mentioned in the introduction,
these results are given in the specific kinematic domain
of both experiments,
at low $x_{Bj}$ ($x_{Bj} < 0.01$) but they take advantage of the large range in $Q^2$, offered by the
HERA kinematics, which covers more than 2 orders
of magnitude. It makes possible to study the transition from
the low $Q^2$ non-perturbative region (around $1$ GeV$^2$) towards higher values of $Q^2$ where the higher twists
effects are lowered (above $10$ GeV$^2$).
The last data on DVCS cross section as a function of $W \simeq \sqrt{Q^2/x}$ are presented on figure \ref{fig1}.
They show a strong $W$ dependence with $\sigma_{DVCS} \propto W^{0.7}$,  characteristic of
a hard process, which can thus be described in perturbative QCD as described in previous sections.

\begin{figure}[!htbp]
\vspace*{-1cm}
 \includegraphics[totalheight=6.2cm]{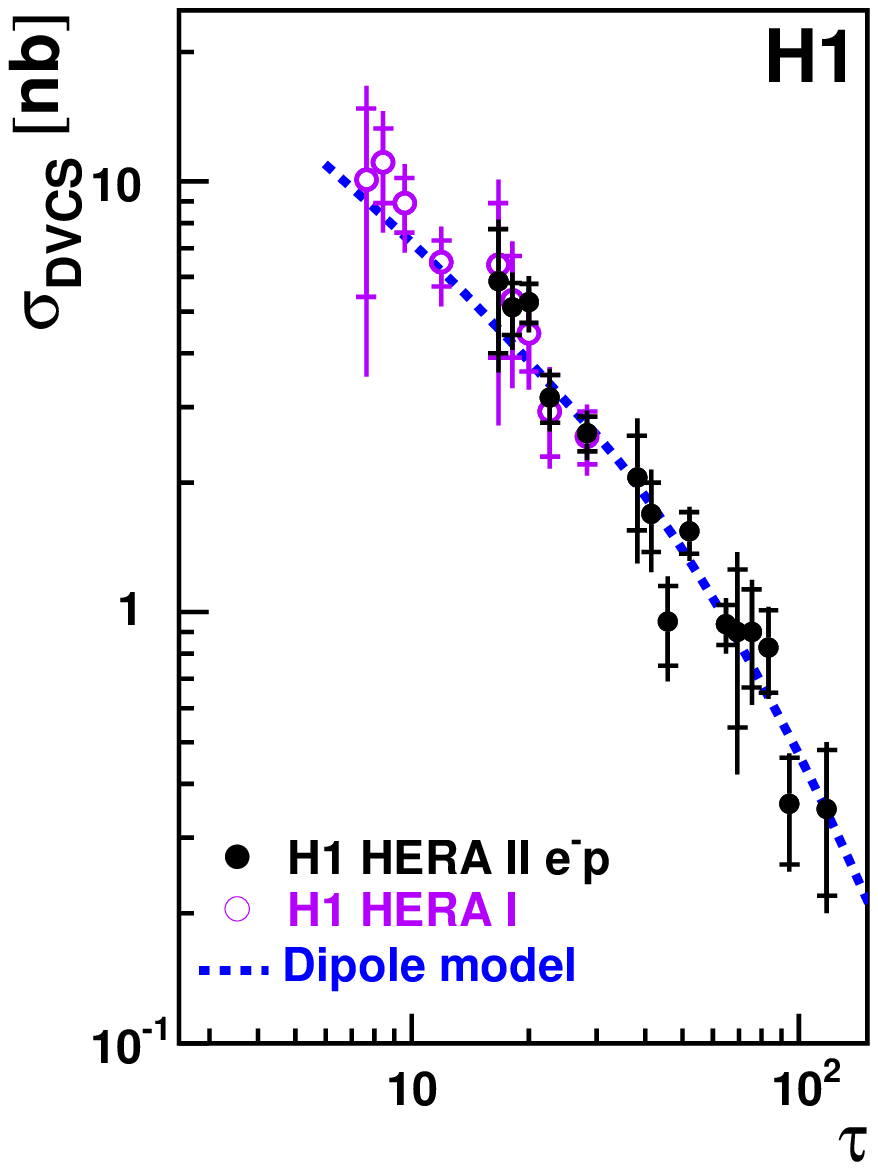}
 \includegraphics[totalheight=6.2cm]{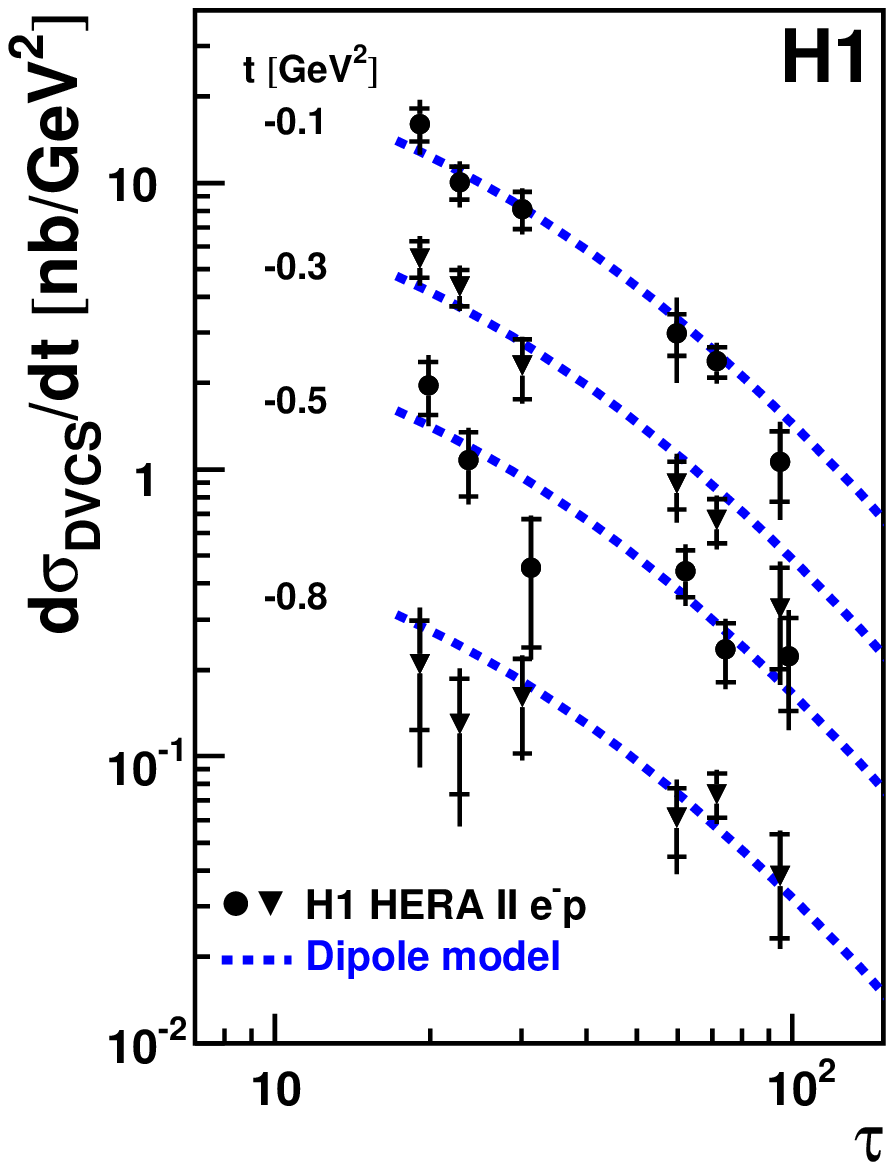}
\caption{\label{figdipole} 
DVCS cross section measurements as a function of 
the scaling variable
$\tau={Q^2}/{Q_s^2(x)}$. 
Results are shown
for the full $t$ range  $|t| <$ 1 GeV$^2$ (left)
and at four values of $t$ (right).
The dashed curves represent the predictions of the 
dipole model \cite{marquet}.
}
\end{figure}

\begin{figure}[ht]
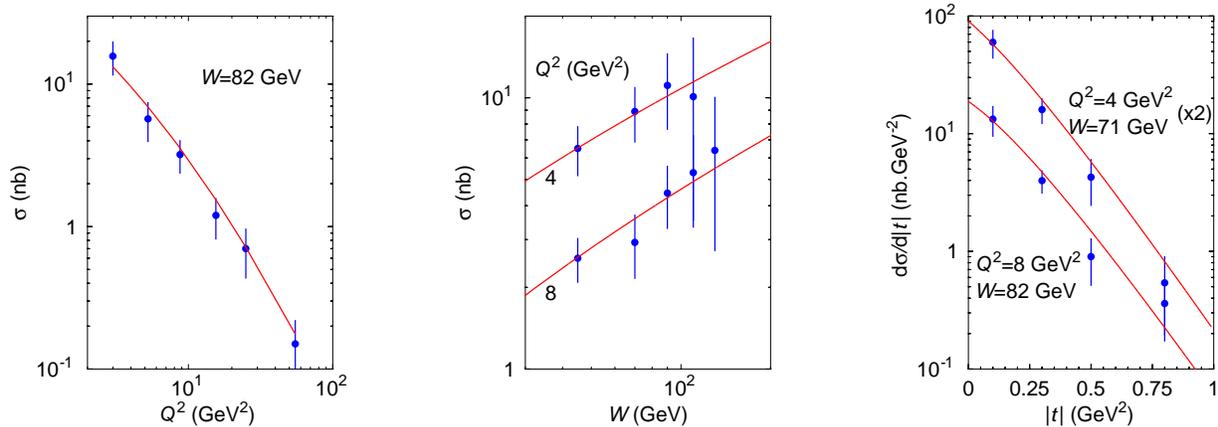

  \includegraphics[scale=0.65]{sigma_dvcs0.ps} \hspace{1cm}
  \includegraphics[scale=0.65]{sigma_dvcs.ps} \hspace{1cm}
  \includegraphics[scale=0.65]{dsdt_dvcs.ps}
\caption{Predictions for the DVCS cross section \cite{robi} with H1 data.}
\label{fig:dvcs}
\end{figure}

Data and model comparisons are presented on figures \ref{figdipole} and \ref{fig:dvcs}.
They show that the dipole approach is 
very efficient in describing the DVCS measurements for the HERA kinematics.
It must be noticed here that the non-forward kinematics for DVCS is encoded only 
through the different weights coming from the
photon wavefunctions in Eq. (\ref{dvcsdip}). It means that without taking
into account non-diagonal effects that drive all the physics of GPDs, we can get
a good descritption of all existing low $x_{Bj}$ data in the framework of the dipole model.

\begin{figure}[!]
    \includegraphics[width=7cm]{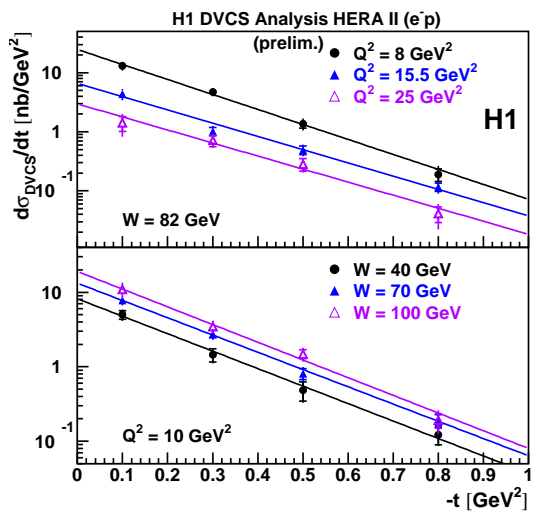}
  \caption{DVCS cross section, differential in $t$ presented with
              a fit of the form $e^{-b|t|}$. 
}
\label{fig1b}  
\end{figure}

\begin{figure}[!] 
\vspace{-2cm}
    \includegraphics[width=6.5cm]{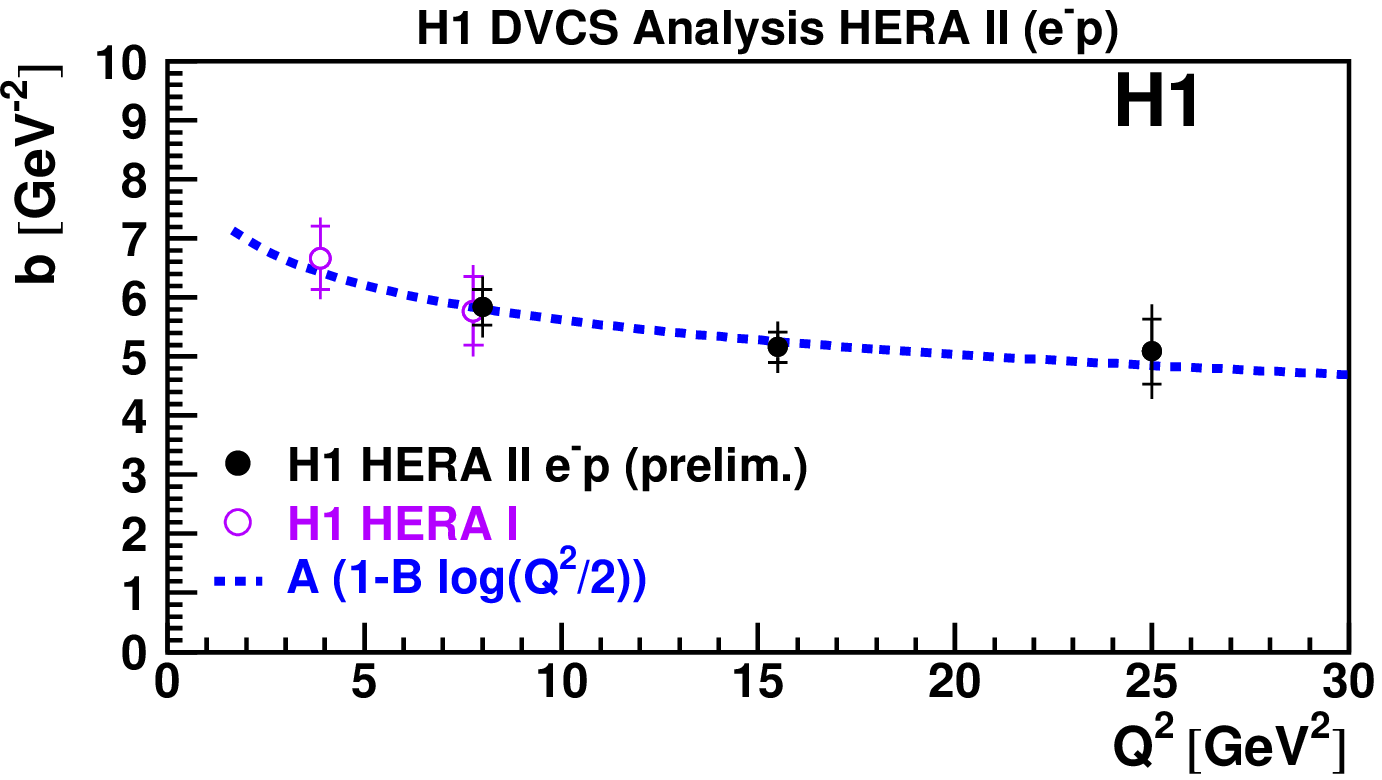}
    \includegraphics[width=6.5cm]{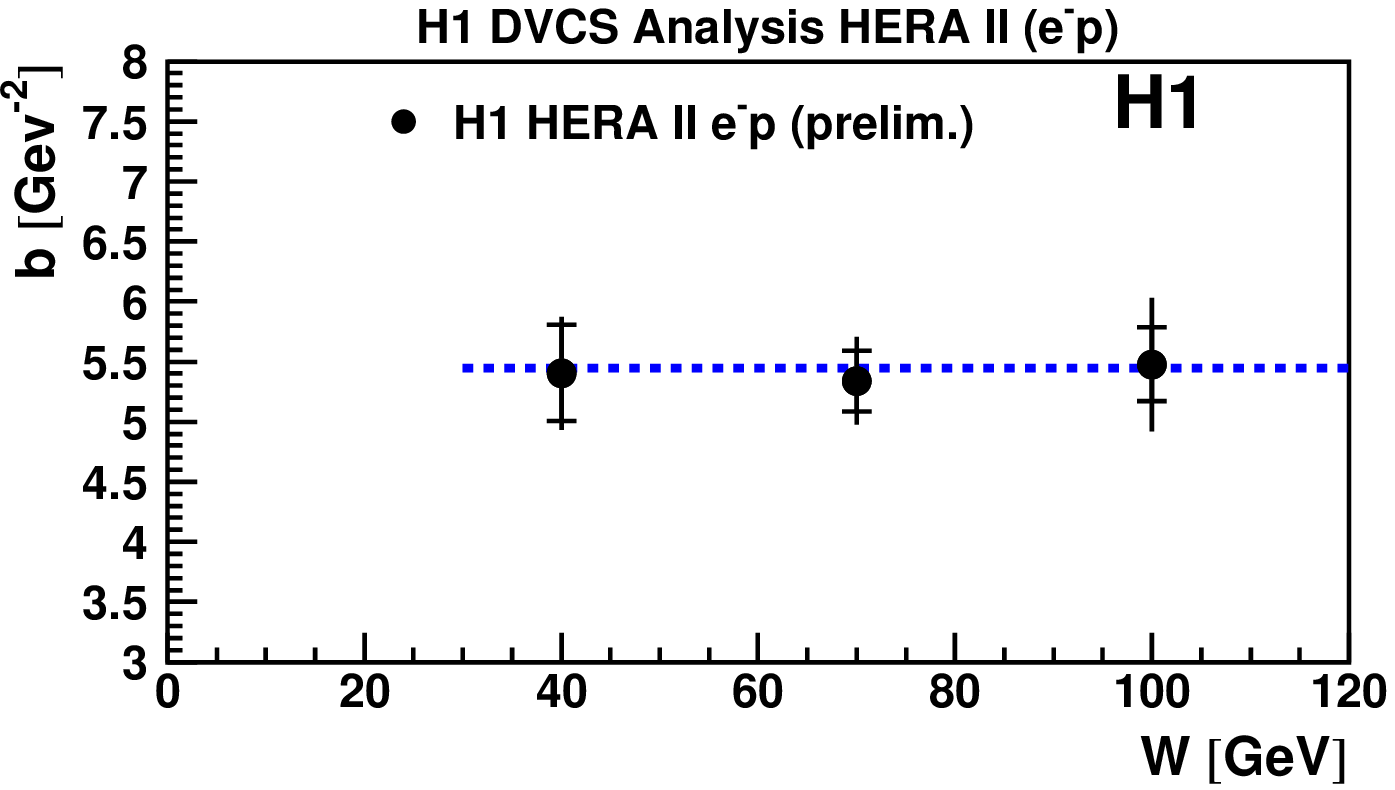}
  \caption{The logarithmic slope of the $t$ dependence
  for DVCS exclusive production, $b$ as a function of $Q^2$ and $W$, extracted from a fit
  $d\sigma/dt \propto
\exp(-b|t|)$  where $t=(p-p')^2$.
}
\label{fig2}  
\end{figure}

Beside $W$ and $Q^2$ DVCS cross sections, a major experimental achievement of H1 \cite{dvcsh1} has been the measurement of
DVCS cross sections, differential in $t=(p'-p)^2$, the momentum transfer (squared) at the proton vertex.
Some results are presente on figure \ref{fig1b}: we observe  the good description
of $d\sigma_{DVCS}/dt$ by a fit of the form $e^{-b|t|}$. Hence, an extraction of the $t$-slope parameter $b$ is accessible
for different values of $Q^2$ and $W$ (see figures \ref{fig1b} and \ref{fig2}). We exploit these values of $b$ in a 
section below.

\section{Quantifying  skewing effects}

Let's define a basic quantity giving an overall measurement of the skewing
properties \cite{lolocompass,lolocompass2}, which includes both the non-forward kinematics and the
possible non-diagonal effects. Namely, we set the ratio between the imaginary parts
of the DIS and DVCS (forward) scattering amplitudes at zero momentum
transfer:
\beq
R \equiv \frac{Im\, {\cal  A}\,(\gamma^*+p \to \gamma^* +p)|_{t=0}}
{Im \,{\cal A}\,(\gamma^*+p \to \gamma +p)|_{t=0}} \ ,
\label{R_def}
\eeq
where $t$ is the square of the four-momentum exchanged at the proton vertex.

\begin{figure}
  \epsfig{figure=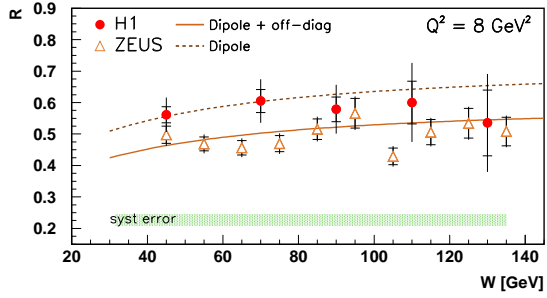,width=0.5\textwidth}
 \vspace*{-0.5cm}
 \caption{\sl The skewing factor $R$ as a function of $W$ at $Q^2=8$~GeV$^2$. 
  The curves represent the theoretical predictions of the  dipole
  approach (see
  text).
}
 \label{fig:r_dip_w}
\end{figure}

In \cite{lolocompass,lolocompass2}, it has been shown how to extract this quantity
from  the  DVCS cross sections.
The factor $R$ is presented  as a function of energy $W$ in
figure ~\ref{fig:r_dip_w}. An almost flat $W$ dependence is observed
within the present precision. This feature can be easily understood  since the $W$ dependence of both the DIS and
DVCS cross section is power-like having a proportional effective power.
Namely, $\sigma_{\mathrm{DIS}}\propto W^{2\lambda}$ and
$\sigma_{\mathrm{DVCS}}\propto W^{4\lambda}$. The mean value $R\simeq 0.5$
is consistent with its early theoretical estimates using the aligned jet
model  and the colour dipole  model (see figure \ref{fig:r_dip_w}).
We also display a dipole model prediction with an 'ad hoc' parametrisation for the
off-diagonal effects. It does not improve the quality of the description of data within the 
present uncertainties.

\section{Nucleon Tomography}

Measurements of the $t$-slope parameters $b$
are key measurements for almost all exclusive processes,
in particular DVCS. Indeed,
a Fourier transform from momentum
to impact parameter space readily shows that the $t$-slope $b$ is related to the
typical transverse distance between the colliding objects \cite{buk}.
At high scale, the $q\bar{q}$ dipole is almost
point-like, and the $t$ dependence of the cross section is given by the transverse extension 
of the gluons (or sea quarks) in the  proton for a given $x_{Bj}$ range.
More precisely, from the  generalised parton distribution  defined in the introduction, we can compute
a parton density which also depends on a spatial degree of freedom, the transverse size (or impact parameter), labeled $R_\perp$,
in the proton. Both functions are related by a Fourier transform 
$$
PDF (x, R_\perp; Q^2) 
\;\; \equiv \;\; \int \frac{d^2 \Delta_\perp}{(2 \pi)^2}
\; e^{i ({\Delta}_\perp {R_\perp})}
\; GPD (x, t = -{\Delta}_\perp^2; Q^2).
$$
Thus, the transverse extension $\langle r_T^2 \rangle$
 of gluons (or sea quarks) in the proton can be written as
$$
\langle r_T^2 \rangle
\;\; \equiv \;\; \frac{\int d^2 R_\perp \; PDF(x, R_\perp) \; R_\perp^2}
{\int d^2 R_\perp \; PDF(x, R_\perp)} 
\;\; = \;\; 4 \; \frac{\partial}{\partial t}
\left[ \frac{GPD (x, t)}{GPD (x, 0)} \right]_{t = 0} = 2 b
$$
where $b$ is the exponential $t$-slope.
Measurements of  $b$
presented in figure \ref{fig2}
corresponds to $\sqrt{r_T^2} = 0.65 \pm 0.02$~fm at large scale $Q^2$ for $x_{Bj} < 10^{-2}$.
This value is smaller that the size of a single proton, and, in contrast to hadron-hadron scattering, it does not expand as energy $W$ increases.
This result is consistent with perturbative QCD calculations in terms of a radiation cloud of gluons and quarks
emitted around the incoming virtual photon. In short, gluons are located at the preiphery of the proton as measured 
here and valence quarks are assumed to form the core of the proton at small value of $\sqrt{r_T^2}$.

In other words,  the
  Fourier transform of the DVCS amplitude is the amplitude to find quarks at
$R_\perp$ in an image plane after focusing by an idealized lens.
The square of the profile amplitude, producing the PDF (in transverse plane)
is positive, real-valued, and corresponds to the
image, a weighted probability to find quarks in the transverse
image plane.  

\section{Perspectives at CERN}
The complete parton imaging in the nucleon would need to get  measurements of $b$ for
several values of $x_{Bj}$, from the low $x_{Bj} < 0.01$ till $x_{Bj}>0.1$. Experimentally,
it appears to be impossible. Is it the breakout of quark and gluon imaging in the proton?
In fact, there is one way to recover $x_{Bj}$ and $t$ correlations over the whole $x_{Bj}$
domain: we need to measure a Beam Charge Asymmetry (BCA) \cite{lolopic,lolocompass,lolobca}.

\begin{figure}[htbp] 
    \includegraphics[width=6cm]{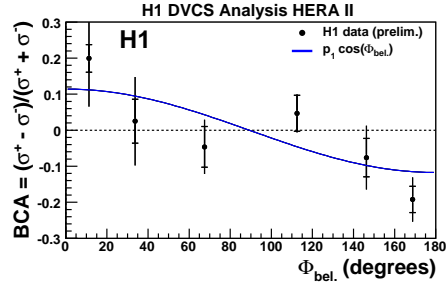}
  \caption{Beam charge asymmetry as a function of $\phi$ measured by H1 \cite{lolopic,lolobca}.
}
\label{fig3}  
\end{figure}

A determination of a cross section asymmetry with respect to the beam
charge has been realised by the H1 experiment by measuring the ratio
$(d\sigma^+ -d\sigma^-)/ (d\sigma^+ + d\sigma^-)$ as a function of $\phi$,
where $\phi$ is the azimuthal angle between leptons and proton plane \cite{lolopic,freund2}.
The result is presented on figure \ref{fig3} with  a fit in $\cos \phi$.
After applying a deconvolution method to account for the  resolution on $\phi$,
the coefficient of the $\cos \phi$ dependence is found to be $p_1 = 0.17 \pm 0.03 (stat.) \pm 0.05 (sys.)$.
This result represents obviously a major progress in the understanding of the very recent field of the 
parton imaging in the proton. We are at the hedge of the giving a new reading on the most fundamental question to know
how the proton is built up by quarks and gluons.

Feasabilities for future BCA measurements at COMPASS have been studied extensively
in the last decade \cite{dhose}. COMPASS is a fixed target experiment which can use
100 GeV muon beams and hydrogen targets, and then access experimentaly the DVCS process $\mu p \rightarrow \mu \gamma p$.
The BCA can be determined when using positive and negative muon beams.
One major interest is the kinematic coverage from $2$ GeV$^2$ till $6$ GeV$^2$ in $Q^2$
and  $x_{Bj}$ ranging from $0.05$ till $0.1$. It means that it is possible to avoid
the kinematic domain dominated by higher-twists and non-perturbative effects 
(for $Q^2 < 1$ GeV$^2$) and keeping a
$x_{Bj}$ range which is extending the HERA (H1/ZEUS) domain.

\section{Summary}
DVCS measurements in the HERA kinematics at low $x_{Bj}$ are well described within a dipole approach,
which encodes
the non-forward kinematics for DVCS only 
through the different weights coming from the
photon wavefunctions. Note that the dipole model presents the great advantage
to define a unique framework that gives a good description of all  hard processes
accessible at HERA. For the first time, we have also shown that proton tomography
enters into the experimental domain of high energy physics, with a first experimental evidence
that gluons are located at the periphery of the proton. A new frontier in understanding
this structure would be possible at CERN within the COMPASS experimental setup.

\end{document}